\documentclass[psf,epsf,twocolumn,showpacs,preprintnumbers,floatfix,prb,superscriptaddress]{revtex4}
\usepackage{graphicx}
\usepackage{graphicx}
\usepackage{dcolumn} 
\usepackage{bm}
\usepackage{epsfig}
\begin{document} 

\title{Compensated magnetism by design in double perovskite oxides}

\author{Victor Pardo}
\email{victor.pardo@usc.es}
\affiliation{Department of Physics, University of California, Davis, CA 95616}
\affiliation{Departamento de F\'{i}sica Aplicada, Facultad de F\'{i}sica, Universidad de Santiago de Compostela, E-15782 Campus Sur s/n, Santiago de Compostela, Spain}

\author{Warren E. Pickett}
\affiliation{Department of Physics, University of California, Davis, CA 95616}

\date{\today}

\begin{abstract}
Taking into account Goodenough's superexchange rules, including both full structural relaxation and
spin-orbit coupling, and checking strong correlation effects, we look for compensated half metals within
the class of oxide double perovskites materials.   Identifying likely half metallic 
(or half semimetallic) antiferromagnets, the full complications including orbital magnetism are
included in order to arrive at realistic predictions of designed magnetic compounds with (near)
vanishing net moment.  After sorting through
several candidates that have not been considered previously, two materials,
K$_2$MnRhO$_6$ and La$_2$CrWO$_6$, remain as viable candidates. An important factor is obtaining
compounds either with very small induced orbital moment (helped by closed subshells) or with an orbital
moment that compensates the spin-orbit driven degradation  of half metallic character. 
While thermodynamic stability of these materials cannot be
ensured, the development
of layer-by-layer oxide deposition techniques does not require that materials be thermodynamically stable
to be synthesized.
\end{abstract}
\maketitle

\section{Introduction}

Double perovskites have drawn considerable attention in the field of spintronics, particularly 
since the discovery\cite{nature_srfemo} of colossal magnetoresistance in Sr$_2$FeMoO$_6$ that 
is believed to be a half metallic (HM) ferromagnet (FM). The HM property is highly desired in spintronics
applications, where spin currents are utilized as well as charge currents.  The double perovskite
family of oxides has been one of the most popular classes within which to look for HMFMs.\cite{katsnelsonRMP}

Recently, with wider application  
of deposition techniques leading to better materials and improved understanding, layer by layer 
deposition of perovskite oxides has led to unusual and potentially useful physical properties.\cite{chakalian}
Most such depositions of perovskites have used the (001) growth direction, resulting in a wide
variety of oxide heterostructures with tetragonal symmetry.
Double perovskite materials can be considered as single unit cell, multilayered perovskite structures 
that have been grown along the (111) direction, and such growth is a promising direction for synthesizing
new members of this class.

Although HMFM character will provide new (and highly sought after) magnetoelectronic
properties, there is a subclass that is more exotic.  When half metallicity occurs and in addition the moments 
compensate, a ``compensated half metal'' (historically called a half metallic antiferromagnet\cite{leuken}) arises.
This state has magnetic order but vanishing macroscopic spin magnetization, it could support a new
type of superconductivity,\cite{sss} and one can imagine 
numerous possible applications for a half metal that is relatively impervious to external magnetic 
fields.  However, this spin-only picture is an idealization; spin-orbit coupling (SOC) couples the two
spin directions and thereby destroys the precise spin
compensation, and in addition generates spin-induced orbital moments.  Thus a fully compensated half 
metal with vanishing macroscopic magnetization is an idealization, and the focus must be on finding
realizations that are as close as possible to full compensation.  

Interest has been rekindled
recently in the effects of spin-orbit coupling in oxides because,
unlike the degradation of half metallicity that is commonly discussed
and expected,
it can also lead to more exotic effects in the spin- and orbital-magnetism,
especially in
$d^1$ or d$^5$ ions in $t_{2g}$ subshells where large orbital moments may arise.\cite{eschrig}
A dramatic possibility is the promotion of compensating spin and
orbital moments.
In Ba$_2$NaOsO$_6$, for example, calculations\cite{bnoo} indicate that
the $d^1$ spin moment of Os is compensated by the $t_{2g}$ L=1 orbital
moment induced by SOC, and it is the partial orbital quenching by the
environment that destroys the compensation of the moment.
In Sr$_2$CrOsO$_6$, neglect of SOC leads to a semimetallic compensated
half metal, which is destroyed by the large SOC in this system.\cite{scoo}
In isovalent Sr$_2$CrRuO$_6$, where weaker SOC arises on Ru compared
with  Os, the reduced spin and induced orbital moments
conspire to nearly cancel, giving
a real possibility for an effectively compensated half metal.  In
correlated insulators, strong SOC can lead to much more complex
magnetic coupling, thereby complicating the resulting magnetic
order\cite{jackeli} and reducing the likelihood of achieving a
compensated half metal.
   
Half-metallic antiferromagnetic (HMAF) materials have not been obtained experimentally yet.  Several 
theoretical efforts have been carried out in predicting half-metallic antiferromagnetism, but experimentally, 
this state remains elusive. HMAFs have been predicted for various double 
perovskites,\cite{pickett_hmafm,lasrvruo6,lacavoso6,lacavmoo6} tetrahedrally coordinated 
compounds\cite{tetra} and for Heusler-structured materials.\cite{cr2mnz,heusler1,heusler2}
Design of materials with desired properties is a long-standing hope that is gaining
momentum,\cite{canfield} and the considerable experience that has been accumulated
in oxide double perovskites provides guidelines to focus design efforts.

In this paper we build on previous studies, and extend the earlier work by considering all of factors: ion
size, structural relaxation, and SOC.
Our general strategy is:\\
i) we focus on the double perovskite structure, with the chemical formula A$_2$MM'O$_6$, where 
M and M' are transition metal ions lying on a rocksalt type sublattice. 
An ordered double perovskite is more likely to occur if M and 
M' differ in ionic radius, and differ in formal charge.\\
ii) we use ab initio calculations to relax the structure, including volume optimization, $c/a$ 
optimization, and internal coordinate relaxation, thereby obtaining structures that are
dynamically stable, and energetically metastable if not stable.\\
iii) we include SOC, which mixes the spin moment with an orbital moment,
 and also analyze the effects of correlation corrections by using the LSDA+U 
method,\cite{lsdau1,lsdau2} to check 
the dependence of the magnetic and electronic structure properties on U 
(the on-site Coulomb 
repulsion).  The objective is to determine to what degree the spin-compensation 
is maintained, and to understand the interplay of structural changes, SOC, and effects of electronic
correlation.

We use the Goodenough-Kanamori-Anderson (GKA) rules for superexchange\cite{goody} for 
the 180$^{\circ}$ metal-oxygen-metal bond occurring in perovskites to guide our choices. 
According to these authors, and the expected moments, only a few 
d$^n$-d$^m$ electronic configurations that would lead to a spin-compensated 
situation are likely. 
In addition, we aim to maximize the likelihood of obtaining 
an ordered perovskite by mixing a 3d row element with another metal cation from 
the 4d or 5d row, so that the size mismatch will make chemical ordering 
more likely. According to GKA, the possible antiferromagnetically coupled electronic 
configurations are: d$^2$-d$^2$, d$^2$-d$^4$(LS), d$^3$-d$^3$, d$^8$-d$^8$ 
and other combinations that include in all cases high spin (HS) cations: 
d$^4$-d$^4$, d$^5$-d$^5$, d$^4$-d$^6$, d$^6$-d$^6$, d$^7$-d$^7$. These 
latter options have not been pursued because it is difficult to stabilize 
HS states with those large moments for the 4d and 5d elements,  which
usually take on a higher valence state than the 3d element, and hence the 
relatively strong crystal field will favor low-spin states. Particularly 
interesting is the d$^3$-d$^3$ configuration when SOC is 
taken into account. Since it deals with a completely filled 
t$_{2g}$ band, the effective angular momentum will be very small, although there is
also the possibility of insulating states. 
We will discuss this issue later in the paper.

\section{Computational details}

Electronic structure calculations were  performed within density functional 
theory\cite{dft} using {\sc wien2k},\cite{wien} which utilizes an augmented 
plane wave plus local orbitals (APW+lo)\cite{sjo} method to solve the Kohn-Sham 
equations. This method uses an all-electron, full-potential scheme that makes no shape 
approximation to the potential or the electron density. The exchange-correlation 
potential used was the Perdew and Wang version of the local density 
approximation\cite{lda} and strong correlation effects were introduced by means 
of the LSDA+U scheme\cite{sic,ylvisaker} including an on-site U (Coulomb repulsion) 
and J (Hund's rule exchange) for the metal d states.
We have used option nldau= 1 in {\sc wien2k}, i.e. 
the so-called ``fully-localized limit". Structural optimizations (unless otherwise stated) were
performed using the Perdew-Burke-Ernzerhof version of the generalized gradient approximation.\cite{gga}
Spin-orbit coupling was introduced using the scalar relativistic approximation.
All the calculations were converged with respect to all the parameters
used, to the precision necessary to support our calculations (converged forces and total energy 
differences), up to R$_{mt}$K$_{max}$= 7 and an 8 $\times$ 8 $\times$ 8 k-mesh.

\section{Revisiting previously found materials}

\subsection{La$_2$VCuO$_6$: d$^1$-d$^9$}

\begin{figure}[ht]
\begin{center} 
\includegraphics[draft=false,width=\columnwidth]{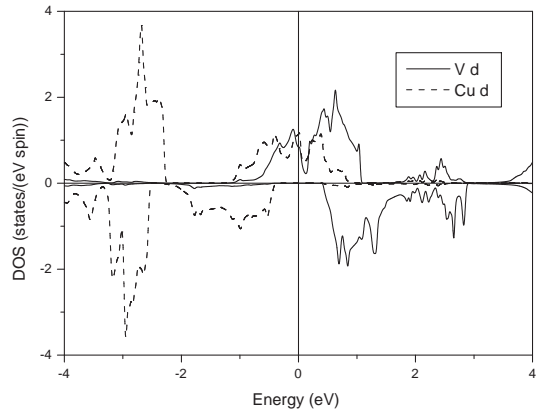}
\end{center} 
\caption{Density of states plots of the tetragonal double perovskite compound La$_2$VCuO$_6$
in a HMAF state within an LSDA+U scheme with U= 3.5 eV for both V and Cu. Upper (lower) panel
shows up (down) spin channel. The spin moment is precisely compensated.}
\label{la2vcu}
\end{figure}

It had been reported previously that this material is a HMAF,\cite{pickett_hmafm,la2vcuo6} but previous studies 
did not consider the structural relaxation that we have now carried out, 
and that is expected to be important in this strongly Jahn-Teller active 
system.
At the LSDA+U level, this compound converges to a HMAF, which remains 
when one relaxes the structure. We have done a relaxation within a 
tetragonal symmetry (but the ratio $c/a$ $\sim$  1.00, to 1\% accuracy, as expected), obtaining 
$a$= 3.64 \AA. The internal structure relaxes to accommodate the Jahn-Teller distortion 
of the Cu$^{2+}$ and V$^{4+}$ cations. Depending on the value of U chosen 
(the same U was chosen for both cations in our calculations), this distortion 
varies, being larger for a larger U. For relatively small values of U 
(effective U of only 4 eV) there is a large Jahn-Teller distortion, the 
oxygen octahedra around the Cu$^{2+}$ cation become elongated by 
about 4\% to accommodate the d$^9$ cation and around V$^{4+}$, the octahedron 
is shortened along one axis by about 4\% to accomodate a d$^1$ cation.
The system becomes a Mott insulator for U= 4.8 eV and J= 0.7 eV.
The electronic structure is presented in 
Fig. \ref{la2vcu} for U= 3.5 eV, where the system is still half-metallic.
 
Figure \ref{la2vcu} shows that half-metallicity occurs with the Fermi level falling
within a 2 eV wide band of Cu e$_g$ and  V t$_{2g}$ character, even when degeneracy
is broken by the Jahn-Teller distortion.  The bottom of the band (a two band complex)
is strongly Cu d in character, while the V d character dominates the upper part of the bands.

We expect a small influence of SOC because only 3d elements are involved. The 
full moment compensation that occurs within LSDA+U will be kept when 
including SOC. 
This d$^9$-d$^1$ 
configuration has small magnetic coupling according to GKA rules (AF 
according to our calculations), and is spin-compensated according to our results. The orbital angular 
momentum of the Cu atoms is 0.1 $\mu_B$, but is negligible for V and the electronic 
structure is not affected by the introduction of SOC. Spin-compensation remains 
unaltered when SOC is introduced, thus the total moment of the system 
is approximately 0.1 $\mu_B$/f.u., quite close to the desired result.

\subsection{Sr$_2$NiOsO$_6$: d$^8$-d$^2$}

This material has been synthesized\cite{srnios_expt} recently and also has been studied using 
ab initio methods.\cite{srnios_abinit} Experimental results were interpreted in terms of a 
competition between FM and AF interactions, caused by the d$^8$-d$^2$ (Ni$^{2+}$-Os$^{6+}$)
electronic structure.  The closed subshell (d$^8$), open subshell (d$^2$) should favor
half metallicity.
According to GKA rules, the e$_{g}^2$-O-e$_{g}^0$ FM superexchange should be the strongest in this system. However, the magnetic susceptibility 
data shows both a positive $\Theta_{CW}$ (showing that FM is dominant at high temperature) and an AF downturn below 50 K, indicating that AF correlations exist at low temperature. Our ab initio calculations show the FM and AF configurations 
are similar in total energy, but FM is distinctly the ground state by 20 meV/metal atom (using
the experimental structure). Also, the spin-compensation that results within LSDA+U is degraded
seriously by SOC, with the large orbital angular momentum leading to a total moment of approximately 
1 $\mu_B$/f.u.  This value arises from orbital moments of
0.5 $\mu_B$ for Os and 0.2 $\mu_B$ for Ni, plus the 0.2 $\mu_B$ from the total spin moment of the system 
arising from SOC-induced spin mixing.
Hence, the condition of a small magnetic moment is not obtained for this compound. The d$^2$-d$^8$ 
electronic structure has the drawback that, when an antialigned spin configuration is set, the 
orbital angular momenta align and add up rather than (partially) canceling (Hund's third rule
for the atoms).

We have performed a structural analysis similar to the one described for La$_2$VCuO$_6$ with the 
same preconditions. Comparing with experimental structural data,\cite{srnios_expt} 
our calculated volume is quite good (1\% larger than experiment) and the $c/a$ ratio is 1\%
smaller than the experimental value.

\begin{table}[h!]
\caption{Calculated versus experimental structural parameters and selected bond lengths for 
Sr$_2$NiOsO$_6$ showing agreement to within 1\% accuracy in the lattice constants.}\label{tab_srnios}
\begin{center}
\begin{tabular}{|c|c|c|}
\hline
Parameter & Experimental\cite{srnios_expt} & This work \\
\hline
\hline
$a$ & 5.49 & 5.53  \\
c & 7.99 & 7.94  \\
Ni-O distance ($\times$ 4) & 2.02 & 1.99 \\
Ni-O distance ($\times$ 2) & 2.04  & 2.03  \\
Os-O distance ($\times$ 4) & 1.91 & 1.92 \\
Os-O distance ($\times$ 2) & 1.96 & 1.94 \\
\hline
\end{tabular}
\end{center} 
\end{table}

\begin{figure}
\begin{center}
\includegraphics[draft=false,width=\columnwidth]{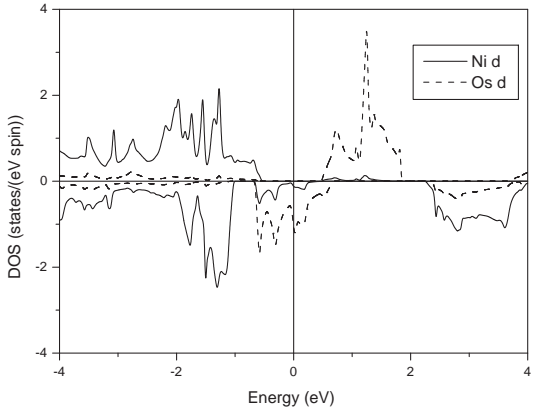}
\end{center} 
\caption{Density of states plots of the tetragonal double perovskite Sr$_2$NiOsO$_6$
in a HMAF state within an LSDA+U scheme with U= 5 eV for Ni and 3 eV for Os, including spin-orbit coupling. Upper (lower) panel
shows up (down) spin channel.}\label{sr2nios}
\end{figure}

For large values of U this compound becomes a Mott insulator. Half-metallicity occurs only for U 
smaller than 3.5 eV for Os (with J= 0.7 eV) and values of U not larger than 7 eV for Ni. 
Realistic values of U and J would be close to the metal-insulator transition. No experimental 
information on its conduction properties is available so far. For smaller values of U, we obtain 
conduction coming from the Os majority spin channel (the unfilled subshell), 
as can be seen in Fig. \ref{sr2nios}.

\section{New findings}

Two materials with the desired properties have been clearly identified. Both possess a d$^3$-d$^3$ 
electronic configuration so a high-spin, magnetic band insulator situation is a possibility. This 
particular electronic configuration has several advantages, that we will discuss, over other 
possible electronic configurations that will promote a spin-compensated state, and they have AF 
coupling by the GKA rules, so they become plausible candidates for being spin-compensated 
half metals or half semimetals. In addition, we have analyzed a compound with a 
d$^2$-d$^2$ configuration, and we
discuss distinctions.

\subsection{Sr$_2$VReO$_6$: d$^2$-d$^2$}

This compound is calculated to have a d$^2$-d$^2$ electronic configuration.
The unoccupied t$_{2g}$ shells lead to a distortion from cubic symmetry. 
Relaxing the structure within tetragonal symmetry, the resulting $c/a$ ratio is 1.02, 
together with a sizable elongation of the oxygen cages to accommodate the d$^2$ electronic state 
(3\% difference in the V-O distance and a smaller 1\% difference for the Re-O distance). 
A summary of the structural parameters obtained is
given in Table \ref{tab_srvre}.

\begin{table}[h!]
\caption{Calculated structural parameters for Sr$_2$VReO$_6$ in the space group I4/mmm (no. 139). $a$= 5.52 \AA ($\sqrt{2}$ $\times$ 3.90 \AA), c=7.96 \AA (2 $\times$ 3.98 \AA).}\label{tab_srvre}
\begin{center}
\begin{tabular}{|c|c|c|}
\hline
Atom & Wyckoff position & Atomic parameter \\
\hline
\hline
Sr & 4d &  (0.5,0,0.25) \\
V & 2a &  (0,0,0) \\
Re & 2b &  (0,0,0.5)\\
O1 & 8h &  (0.2521,0.2521,0.5) \\
O2 & 4e &  (0,0,0.2494)\\
\hline
\end{tabular}
\end{center} 
\end{table}

\begin{figure}[ht]
\begin{center} 
\includegraphics[draft=false,width=\columnwidth]{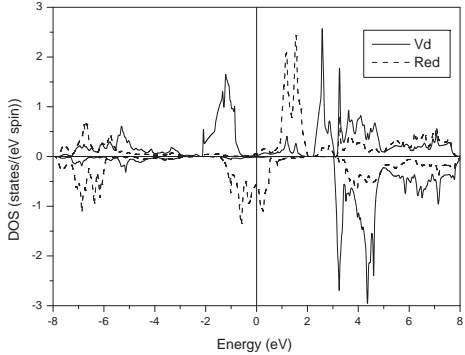}
\end{center} 
\caption{Density of states plots of the tetragonal double perovskite Sr$_2$VReO$_6$
in a HMAF state within an LSDA+U scheme with U= 5 eV for V and 3 eV for Re, including spin-orbit coupling. Upper (lower) panel
shows up (down) spin channel.}\label{sr2vre}
\end{figure}

We have studied the electronic structure of this yet-unreported compound with both LSDA+U and SOC.  
Due to slight band overlap, there is substantial but not total spin-moment compensation, 
leading to a total spin moment of less than 
0.1 $\mu_B$. However, because of the unfilled t$_{2g}$ shells in both cations, large  
orbital angular moments develop, particularly large for Re (0.4 $\mu_B$). The result is a total 
magnetic moment of approximately 0.5 $\mu_B$ for U(V)= 7 eV and U(Re)= 3 eV (J= 0.7 eV for both 
cations). While there is no true gap in the V majority spin channel, there is a very large fractional spin 
polarization in the density of states at the Fermi level, as can be seen in Fig. \ref{sr2vre}.

\subsection{K$_2$MnRhO$_6$: d$^3$-d$^3$}

As we have just seen in the case of Sr$_2$VReO$_6$, having unfilled shells encourages orbital 
angular moments. In that respect, the following two examples are materials which
have a d$^3$-d$^3$ electronic configuration, so the filled t$_{2g}$ shells will give very
small orbital moments to obscure the spin compensation. 

We calculate that this K$_2$MnRhO$_6$ compound has the configuration Mn$^{4+}$: d$^3$ - Rh$^{6+}$: d$^3$, which couples AF. 
The Rh moment is strongly delocalized on the oxygen neighbors (up to 0.3 $\mu_B$ per oxygen 
and only 1 $\mu_B$ in the Rh muffin-tin sphere), but all in all, this compound is near HMAF 
with Fermi level lying very near band edges. The lattice parameter 
$a$ was optimized and atomic positions were relaxed within the Fm$\overline{3}$m space group 
(no. 225), and $a$= 3.94 \AA\ was obtained. The d$^3$-d$^3$ configuration of ions with cubic
symmetry leaves a cubic structure, 
leaving only a small relaxation of the oxygen cages around the cations (Rh-O distance of 2.00 
\AA\, and Mn-O distance of 1.94 \AA).

\begin{figure}[ht]
\begin{center} 
\includegraphics[draft=false,width=0.65\columnwidth]{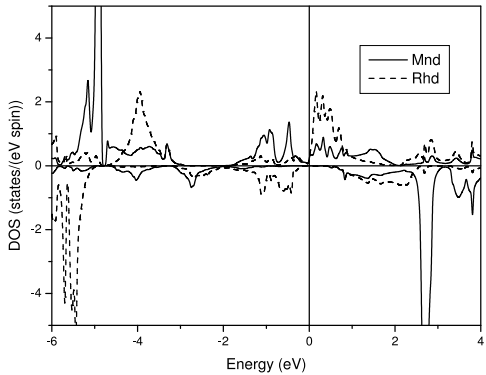}
\end{center} 
\caption{Density of states plots of the tetragonal double perovskite K$_2$MnRhO$_6$
in a HMAF state within an LSDA+U scheme with U= 5 eV for Mn and 3 eV for Rh, including spin-orbit coupling. Upper (lower) panel
shows up (down) spin channel.}\label{k2mnrh}
\end{figure}

As mentioned, very small values of the orbital angular momentum arise when SOC is 
introduced: 0.05 $\mu_B$ on Rh, 0.01 $\mu_B$ on Mn.
Adding spin and orbital angular momenta, the net moment is only 0.02 $\mu_B$, and it 
remains very close to compensated half metal for values of U of 6 eV for Mn and 4 eV for Rh. 
Even though Fig. \ref{k2mnrh} looks similar to a zero-gap situation, on the verge of
a metal-insulator transition, bands cross the Fermi level even for larger values of U, 
and (half) semimetal character is maintained.

\subsection{La$_2$CrWO$_6$: d$^3$-d$^3$}

This is another example of a d$^3$-d$^3$ AF coupling situation, and again, it is a 
not-yet-synthesized material, completely ab initio designed. Structural relaxation was 
carried out allowing tetragonal distortion. 
No tetragonal distortion was obtained, and $a$ = 4.00 \AA. The difference in ionic radii leads 
to a Cr-O distance of 1.97 \AA\ and W-O distance of 2.03 \AA.

\begin{figure}[ht]
\begin{center} 
\includegraphics[draft=false,width=\columnwidth]{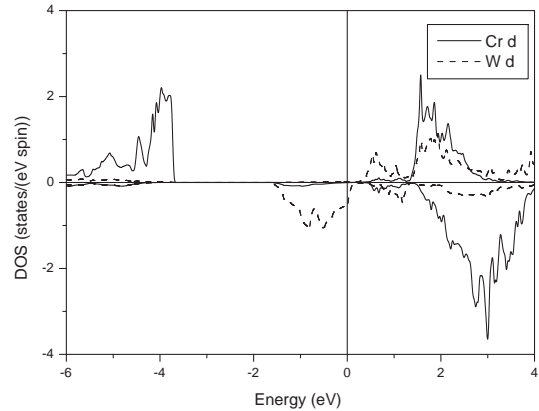}
\end{center} 
\caption{Density of states plots of the tetragonal double perovskite La$_2$CrWO$_6$
in a HMAF state within an LDA+U scheme with U= 5 eV for Cr and 3.5 eV for W, including spin-orbit coupling. Upper (lower) panel
shows up (down) spin channel.}\label{la2crw}
\end{figure}

Evolution with U and SOC indicates that the latter does not seriously degrade the moment 
compensation. For large values of U, where the material is still a metal, the total moment 
stays below 0.2 $\mu_B$/f.u. W has an orbital moment of 0.15 $\mu_B$ and a much smaller one, 
and opposite in an AF configuration, arises from Cr. Even for large values of U (7 eV for Cr 
and 4 eV for W), the system remains metallic (half-metallic) and the total magnetic moment,
including spin and orbital components, does not exceed 0.10 $\mu_B$/f.u. 
Metallicity comes from a broad W d band that crosses the Fermi level, as can be seen in Fig. \ref{la2crw}. An electron pocket is formed in an itinerant W d band and a corresponding La-character hole pocket (not shown) appears.

\section{Unpromising compounds}

On our way to detecting these spin-compensated half-metals, we studied other seemingly promising
double perovskites. We will 
discuss these only briefly, since they do not show the desired behavior.

\subsection{d$^1$-d$^9$}

Even though superexchange rules predict this coupling to be negligible, we have previously 
studied a system that could be promising. Hence, as an analogy with the Jahn-Teller compound 
La$_2$VCuO$_6$, we tried changing each of the 3d elements by a corresponding 4d element, without 
changing the electron count, i.e. La$_2$CuNbO$_6$ and La$_2$VAgO$_6$ which are less likely to
be Mott insulators. In these cases, Nb and Ag respectively are non-magnetic, and the spin-compensation is lost.

\subsection{d$^2$-d$^4$}

This electronic configuration could lead to a spin-compensated state, if the d$^4$ cation is in a low-spin state.
We have tried Ba$_2$FeCrO$_6$ (also with Sr and Ca with similar results), which could assume a d$^2$-d$^4$ electronic configuration. It is indeed the case, but Fe is in a HS state (not allowing for spin compensation to happen). A similar problem appears with the compound K$_2$FeRuO$_6$, where also Fe$^{4+}$ is also in a HS state. This electronic configuration has been studied in the past\cite{lasrvruo6,lacavoso6} for several perovskites with general formula AA'BB'O$_6$.



\subsection{d$^2$-d$^8$}

We have found several cases where always the FM exchange mediated by the e$_g$ electrons is more intense 
than the AF coupling between the t$_{2g}$ electrons. Starting from Sr$_2$NiOsO$_6$ compound discussed above 
and substituting Os by Ru to form Sr$_2$NiRuO$_6$, we find that FM coupling remains. This is the standard 
prediction of GKA rules. Apart from these, we have also identified that FM coupling is more favorable in 
La$_2$CrNiO$_6$ (and also with W and Mo on the Cr site). This material is calculated to be an interesting 
FM insulator (of which there are relatively few examples), where instead of promoting the possible 
Jahn-Teller distorted Ni$^{3+}$ cation, it leads to the formation of a Cr$^{2+}$-Ni$^{4+}$ configuration. 
This charge ordered state is thought to occur to prevent the system from developing a Jahn-Teller distortion 
when it is close to the itinerant electron limit.\cite{CO_vs_JT} 

Disordered LaCr$_{0.5}$Ni$_{0.5}$O$_3$ 
is known experimentally to be insulating,\cite{lacrnio6_mit} but close to a metal-insulator transition 
(the Ni end of the series
LaNiO$_3$ is a metal). Our calculations show that the electronic structure of the charge ordered material 
explains the insulating state found experimentally, together with the prediction of FM coupling between 
the d$^2$ and d$^8$ cations, making the system potentially interesting as another FM insulator. For these 
calculations, we have taken the lattice parameters from the experiment\cite{lacrnio6_struct} with the 
disordered system and relaxed only the internal coordinates. The different ionic radii lead to different 
metal-oxygen bond lengths (1.90 \AA\ for Cr and 1.96 \AA\ for Ni).

\subsection{Antiferrimagnetic insulators}

We have found the following magnetically-compensated insulators with double perovskite structure: i) La$_2$CrMoO$_6$ 
is isoelectronic to the spin-compensated metal La$_2$CrWO$_6$, but introducing Mo instead of W opens up a 
gap at the Fermi level. It is the greater itineracy of the 5d element that leads to the metallic state 
that we analyzed above. ii) In analogy with Sr$_2$CrOsO$_6$,\cite{scoo} we have tried with a bigger 
alkaline-earth cation. The compound Ba$_2$CrOsO$_6$ is found to be an antiferromagnetic insulator, with 
configuration d$^3$-d$^3$. As an isovalent compound, Sr$_2$CrRuO$_6$ is calculated to be an 
antiferromagnetic insulator.

\section{Summary}

In this paper we have suggested two very nearly spin-compensated half metals (or half semimetals): K$_2$MnRhO$_6$ and La$_2$CrWO$_6$, both with a formally d$^3$-d$^3$ configuration. 
It is found that, within a double perovskite structure, a d$^3$-d$^3$ electronic configuration promotes 
the chances to obtain a nearly spin-compensated half-metal, due to several factors.
i) No competition between FM and AF interactions occurs as happens when there is  partial occupation 
of e$_g$ states in one of the metal cations. 
ii) Having the t$_{2g}$ shell completely filled leads to a very small orbital angular moment, which 
works even further in favor of a very small magnetic moment solution. 
iii) Having less than half-filled d shells in both metal ions helps to avoid imbalance in the total moment  
that occurs when spin-orbit coupling is important, specially for 4d and 5d elements. Having both cation 
d shells less than half filled means that the orbital angular momenta of both cations will also orient
antiparallel to the spin, preventing the system from having a larger net magnetic moment. 
iv) The large S=3/2 spins make our calculations more realistic; ordering of low spin ions is often more
affected by quantum fluctuations.  
Both materials that we have discussed at length are designed {\it ab initio}, no experimental information on 
them yet exists to the best of our knowledge.

\section{Acknowledgments}

We have benefited from discussion on this topic with J. Chakhalian.
This project was supported by DOE grant DE-FG02-04ER46111 and through interactions with
the DOE's Predictive
Capability for Strongly Correlated Systems team of the Computational
Materials Science Network. V.P. acknowledges financial support from Xunta de Galicia through the Human Resources Program.

\end{document}